\documentclass[12pt]{article}

\usepackage{amsfonts,,amssymb}

\newcommand{\ra}{\rightarrow}

\newcommand{\ZZ}{{\mathbb Z}}
\newcommand{\RR}{{\mathbb R}}

\newcommand{\tA}{{\tilde A}}

\newcommand{\tF}{{\tilde F}}
\newcommand{\modtwo}{{\ {\rm mod}\ 2}}

\newcommand{\Tr}{{\rm Tr}}

\title{Bosonic Topological Insulators and Paramagnets: a view from cobordisms}

\author{Anton Kapustin \\ {\it California Institute of Technology, Pasadena, CA 91125}}

\begin{document}

\maketitle

\abstract{We classify Bosonic Topological Insulators and Paramagnets in $D\leq 4$ spatial dimensions using the cobordism approach. For $D<4$ we  confirm that the only such phase which does not fit into the group cohomology classification is the 3D Bosonic Topological Insulator protected by time-reversal symmetry whose surface admits an all-fermion topologically ordered state. For $D=4$ there is a unique "beyond group cohomology" phase. It is protected by gravitational anomalies of the boundary theory and is stable without any additional symmetry.

\section{Introduction and summary}

Classification of Symmetry Protected Topological Phases is an important problem which has been addressed using a variety of methods \cite{Kitaev} - \cite{MoradiWen}. The most systematic and general method is the group cohomology approach \cite{GC,GuWen}, but it was shown \cite{VishSent} that it does not cover all possibilities. Recently we proposed a classification scheme for SPT phases based on the assumption that the low-energy effective action is topological and invariant under cobordisms \cite{Kcobord}. This approach can be viewed as a refinement of the group cohomology scheme which properly takes into account gravitational couplings. Within this framework one can write down all possible effective actions in any given dimension  and construct topologically-ordered surface states for each one of them.\footnote{As explained in \cite{Kcobord} there are complications if an SPT phase has a nonzero thermal Hall response. Throughout this paper we only deal with SPT phases whose symmetry group includes time-reversal. Thermal Hall response vanishes for such pahses.}  It was shown in \cite{Kcobord} (see also \cite{Ryan}) that this approach naturally incorporates all known bosonic SPT phases in low dimensions with only time-reversal symmetry $\ZZ^T_2$. The goal of this paper is to study low-dimensional bosonic SPT phases with $U(1)\rtimes \ZZ^T_2$ and $U(1)\times \ZZ^T_2$ symmetries. Following \cite{VishSent}, we will refer to the former class of SPT phases as Bosonic Topological Insulators, while the latter ones will be called Bosonic Topological Paramagnets. The reason for this terminology is that under time-reversal $T$ the electric charge generator $Q$ is even, while the generator of rotations about a fixed axis $J$ is odd. Since $T$ is anti-unitary, we have
$$
T e^{i \alpha Q} T^{-1}=e^{-i\alpha Q},\quad T e^{i \alpha J} T^{-1}=e^{i\alpha J}.
$$

We will denote by $D$ the spatial dimension and by $d$ the space-time dimension, thus $d=D+1$. The results of the classification for $D\leq 4$ are summarized in Table 1. All known bosonic SPT phases with symmetries $U(1)\times  \ZZ_2^T$ and $U(1)\rtimes\ZZ_2^T$ and $D\leq 3$ appear in this table, and no extra phases exist within the cobordism approach. For $D=4$ our results appear to be new. The only phase which does not fit into the group cohomology classification turns out to be protected by gravitational anomalies alone and has already been described in \cite{Kcobord,Ryan}. It is characterized by the property that it admits a gapped surface with a topological $\ZZ_2$ order, fermionic electric excitations, and fermionic vortex strings.

\begin{table}\label{table one}
\begin{center}
\begin{tabular}{c|ccc}
 $D$ & $\ZZ_2^T$ & $U(1)\times\ZZ_2^T$ & $U(1) \rtimes \ZZ_2^T$ \\ \hline
 $0$  &        0        &   $0$                              & $\ZZ$ \\
 $1$ &   $\ZZ_2$    &   $\ZZ_2^2  $                 & $\ZZ_2 $\\
 $2$ &   $0 $           &  $0  $                             & $\ZZ_2$  \\
 $3 $ &  $ \ZZ_2^2$ & $\ZZ_2^4$                   & $\ZZ_2^3$ \\
 $4$ &   $\ZZ_2$     & $\ZZ_2$                    & $\ZZ^2\times\ZZ_2^2$ \\
\end{tabular}
\end{center}
\caption{Classification of bosonic SPT phases with symmetries $\ZZ_2^T$, $U(1)\times\ZZ_2^T$ and $U(1)\rtimes \ZZ_2^T$.}
\end{table}

The organization of the paper is as follows. In sections 2 and 3 we adapt the approach of \cite{Kcobord} to include continuous symmetry groups such as $U(1)$. In the sections 4-8 we classify Bosonic Topological Insulators and Paramagnets for $D\leq 4$. In the appendix we collect some background material on the calculus of simpicial cochains, Stiefel-Whitney classes and Steenrod squares, as well details of some computations.

I am grateful to Ryan Thorngren for discussions and to Ashvin Vishwanath for patiently answering my questions regarding Symmetry Protected Topological Phases. I also would like to acknowledge discussions with Xiao-Gang Wen which helped me to find an error in the first version of the paper. This work was supported in part by the DOE grant  DE-FG02-92ER40701.

\section{Effective action and its properties}

The classification proposal of \cite{Kcobord} was formulated in a way which was convenient for discrete symmetry groups. Namely, it was proposed that for a discrete symmetry group $G$ bosonic SPT phases with symmetry $G$ are classified by the Pontryagin-dual of the torsion subgroup of the twisted bordism group of $BG$, where $BG$ is the classifying space of $G$. The twist is determined by the homomorphism $\rho: G\ra\ZZ_2^T$.  $BG$ can also be thought of as the classifying space of flat $G$-connections, since for discrete $G$ there is no difference between principal $G$-bundles and flat $G$-connections.

When $G$ is continuous, a given principal $G$-bundle typically admits more than one flat $G$-connection. Moreover, a $G$-connection need not be flat, and it is not immediately obvious what the correct version of $BG$ should be.

To circumvent this difficulty, we need to reexamine the physical meaning of the proposal in \cite{Kcobord}. As in the group cohomology approach, the idea is that given an SPT phase with a global symmetry $G$ one can promote $G$ to a gauge symmetry and couple the system to a flat gauge field $A$ for $G$. Then one can integrate out the matter fields and obtain an effective action which depends on $A$ as well as background geometry. The gauge field $A$ must be flat because coupling the system to such a field is unambiguous (this is essentially the minimal coupling prescription which tells us to replace ordinary derivatives with covariant derivatives). In contrast, if the curvature $F=dA+A\wedge A$ is nonzero, one can add to the action terms which depend on $F$, and such terms are far from unique.

SPT phases are usually defined on a spatial lattice, while time may or may not be discretized. In the effective action approach we want to allow space-time to have an arbitrary topology, thus we discretize both space and time and regard the system as being defined on a general triangulation $K$ of a $d$-dimensional manifold $X$. Since in cases of interest $G$ involves time-reversal, $X$ need not be orientable. The lattice analog of a gauge field is a simplicial 1-cochain with values in $G$ (see appendix for a review of the calculus of simplicial cochains). Since $G$ has the form of a (semi)direct product of $U(1)$ and $\ZZ_2^T$, the gauge field can be regarded as a pair of 1-cochains $(w,A)$, where $w$ takes values in $\ZZ_2^T$, and $A$ takes values in $U(1)$.

In what follows it will be important that a flat $U(1)$ gauge field can be a connection on a nontrivial $U(1)$-bundle. In the continuum such a gauge field has a Chern class $c(A)$ which takes values in $H^2(X,\ZZ)$. Its curvature 2-form defines a class in $H^2(X,\RR)$ which vanishes if $A$ is flat. However, $c(A)$ does not necessarily vanish, only its projection to $H^2(X,\RR)$ does. This distinction is important whenever $H^2(X,\ZZ)$ has a torsion part and will be crucial for all our considerations. 

On a triangulated manifold we can define $c(A)$ as follows. We can identify $U(1)$ with the quotient $\RR/\ZZ$. Accordingly, we will represent a $U(1)$ gauge field $A$ by an $\RR$-valued 1-cochain which we will denote $\tA$. Flatness of $A$ then means that $\tF=\delta \tA$ is an integral 2-cochain. This cochain is obviously closed, but in general cannot be written as a differential of an integral 1-cochain. By definition, the cohomology class of $\tF$ is $c(A)$. 

The $\ZZ_2^T$ gauge field $w$ can also be thought of as a 1-cochain. If we regard it as taking values in integers modulo $2$, then it is closed, i.e. $\delta w=0$. One can also regard it as a $\ZZ$-valued 1-cochain, then it is closed modulo $2$. An important difference compared to $A$ is that the class of $w$ in $H^1(X,\ZZ_2)$ is fixed by the manifold $X$. Namely, the holonomy of $w$ must be $1 \modtwo$ along any orientation-reversing closed path and $0 \modtwo$ along any orientation-preserving closed path. In other words, up to a $\ZZ_2$ gauge transformation, $w$ is a representative of the Stiefel-Whitney class $w_1(X)\in H^1(X,\ZZ_2)$. 

We can now describe the lattice data $(w,\tA)$ for the case when $G=U(1)\times\ZZ_2^T$. The gauge field $\tA$ is an $\RR$-valued 1-cochain such that $\tF=\delta \tA$ is an integral 2-cochain, while $w$ is a $\ZZ_2$-valued 1-cochain whose cohomology class is $w_1(X)$. Here the difference between the group cohomology approach and the cobordism approach appears. In the former approach, we would try to classify all effective actions which depend on $w$ and $\tA$ and are invariant under all gauge transformations. In the latter approach, we allow the action to depend on other Stiefel-Whitney classes $w_2(X), w_3(X), \ldots,$ as well. 

The case $G=U(1)\rtimes\ZZ_2^T$ is slightly more involved. Here the gauge field $A$ is a twisted 1-cocycle with values in $\RR^T/\ZZ^T$, where the twist arises from the natural action of $\ZZ_2^T$ on $\RR$ and is determined by $w$.  Thus $\tA$ is now a 1-cochain with values in the local system $\RR^T$ such that $\tF=\delta_w \tA$ is a 2-cocycle with values in the local system $\ZZ^T$. Here $\delta_w$ is the simplicial differential twisted by the 1-cocycle $w$, see appendix for details. Thus $\tF$  defines an element of $H^2(X,\ZZ^T)$ which is the analog of $c(A)$.  Note that since $\ZZ^T/2\simeq \ZZ/2$, $\tF$ modulo $2$ can be regarded as an ordinary (untwisted) 2-cocycle modulo $2$. 

The effective action is assumed to be local, in the sense that it has the form
$$
S=\int_X L,
$$
where $L$ is a $d$-dimensional cocycle built out of $A$ and the Stiefel-Whitney classes. We normalize the action so that the weight in the path-integral is 
$\exp(2\pi i S)$. Thus the action takes values in $\RR/\ZZ$. Integration is really the cap product of the $d$-cocycle $L$ with the fundamental homology cycle of $X$. In the oriented case, the fundamental homology class lives in $H_d(X,\ZZ)$. For our purposes, since $G$ contains orientation-reversing elements, $X$ should be unoriented (and perhaps unorientable). Then the fundamental homology class lives in $H_d(X,\ZZ^T)$. Thus $L$ must be valued in the local system $\RR^T/\ZZ^T$, so that the result of integration is an element of $\RR/\ZZ$.

Note that the local system $\RR^T/\ZZ^T$ has a local subsystem of elements of order $2$. This subsystem is trivial and isomorphic to $\ZZ_2$. If $L$ happens to take values in this subsystem, one can regard $L$ as a $\ZZ_2$-valued $d$-cocycle. Then $S$ can be thought of as the cap product of $L$ with the $\modtwo$ fundamental homology class of $X$. 

\section{Quantum Hall response}

In general the action will depend both on $\tA$ and $\tF=\delta \tA$. If the action depends only on $\tF$, its variational derivative with respect to $A$ vanishes identically, that is, the vacuum current density vanishes in any background. We will say that such an SPT phase has zero quantum Hall response. The quantum Hall response is determined entirely by the terms in the action which depends on $\tA$, not just $\tF$. Such terms are Chern-Simons terms and are well-understood. They exist only for $D=2p$ and do not involve the Stiefel-Whitney classes. On an oriented manifold of dimension $d=2p+1$ the $U(1)$ Chern-Simons action has the form
$$
S_{CS}=n \int_X \tA \tF^p,\quad n\in\ZZ.
$$
Note that we do not have the usual normalization factor $1/(2\pi)^p$ because in our normalization it is $\tF$ which is integral, not $\tF/2\pi$. 

In the unoriented case we must require the integrand to be a $d$-cocycle with values in $\RR^T/\ZZ^T$ rather than $\RR/\ZZ$. For $G=U(1)\times\ZZ_2^T$ both $\tA$ and $\tF$ are ordinary (untwisted) cocycles, and this condition forces $n=0$, that is, the quantum Hall response must vanish. For $G=U(1)\rtimes \ZZ_2^T$ and $p$ odd we have the same result, but for $p$ even this constraint is satisfied for any $n$. Thus Bosonic Topological Insulators may have a nontrivial quantum Hall response for $D=4k$, where $k$ is integral, and possible responses are labeled by an integer $n$.

A related subtlety occurs in dimension $D=4k+1$ and for symmetry group $G=U(1)\rtimes \ZZ_2^T$. There one can write down an action with a continuous parameter $\theta\in \RR/2\pi \ZZ$ which depends only on $\tF$:
\begin{equation}\label{thetaF}
S=\frac{\theta}{2\pi} \int_X \tF^{2k+1}.
\end{equation}
This action makes sense because $\tF$ is a 2-cocycle with values in $\ZZ^T$ and therefore $\tF^{2k+1}$ is a $(4k+2)$-cocycle with values in $\ZZ^T$. Since we classify SPT phases up to homotopy, and the parameter $\theta$ can be continuously deformed to $0$, such an action corresponds to a trivial SPT phase. This has the following implication. For all odd $D$ (i.e. even $d$) and for both $G=U(1)\times \ZZ_2^T$ and $U(1)\rtimes\ZZ_2^T$ we can construct an action which depends only on $F=\tF \modtwo$, namely
\begin{equation}\label{modtwoF}
S=\frac{1}{2}\int_X F^{d/2}.
\end{equation}
But for $d=4k+2$ and $G=U(1)\rtimes \ZZ_2^T$ this action is a special case of the action (\ref{thetaF}) with $\theta=\pi$. Thus this action corresponds to a trivial SPT phase. In all other cases the action (\ref{modtwoF}) corresponds to a nontrivial SPT phase. 

As discussed in \cite{Kcobord}, the action may also depend on the Levi-Civita connection $\omega$ on the tangent bundle of $X$. Terms which depend only on $\omega$ are gravitational Chern-Simons terms and describe thermal Hall response. They are odd under parity and thus must vanish for SPT phases with time-reversal symmetry. But there may also be mixed Chern-Simons terms involving both the $U(1)$ gauge field and $\omega$. Such terms may appear only for $G=U(1)\rtimes \ZZ_2^T$, since only in that case is the gauge field odd under parity. In the range of dimensions we are considering, such a mixed Chern-Simons term exists only for $D=4$. and has the form
\begin{equation}\label{CSmixed}
\frac{n}{(2\pi)^3} \int_X \tA\, \Tr\,\omega^2.
\end{equation}
One physical effect of this term is that Pontryagin density induces electric charge density. Another consequence is a thermal Hall response in the presence of a constant magnetic field.

 \section{$D=0$}

We begin with the case of SPT phases for $D=0$. These are systems with no spatial dimensions, just the time dimension. The action takes values in $\RR/\ZZ$ and has the form
$$
S=\int_X L,
$$
where $L$ is a 1-cocycle with values in $\RR^T/\ZZ^T$. One candidate for $L$ is $w_1(X)$. However, all 1-manifolds are orientable, hence $w_1(X)=0$. 

Another option is to use $A$. In the case $G=U(1)\times\ZZ_2^T$ it takes values in $\RR/\ZZ$ rather than $\RR^T/\ZZ^T$. Hence in this case there are no nontrivial effective actions, and we conclude that there are no Bosonic Topological Paramagnets in $D=0$, in agreement with the group cohomology classification \cite{GC}. Note that although $w_1(X)=0$ and $X$ is orientable, it does not have a canonical orientation, and an action like $\int_X A$ would change sign under orientation-reversal, that is, it would break time-reversal invariance.

On the other hand, for $G=U(1)\rtimes\ZZ_2^T$ $A$ takes values in $\RR^T/\ZZ^T$, just like $L$, hence the most general action is
\begin{equation}\label{CSone}
S=n \int_X A,\quad n\in\ZZ
\end{equation}
Thus Bosonic Topological Insulators in $D=0$ are classified by $\ZZ$. This also agrees with the group cohomology classification \cite{GC}. The integer $n$ has a simple meaning: it is the electric charge of the ground state of the system. It can also be thought of as determining the quantum Hall response (the action (\ref{CSone}) can be thought of as a 1d Chern-Simons action).

\section{$D=1$}

Since the quantum Hall response vanishes for $D=1$, the action must constructed out of $w_1, w_2$ and $\tF$. Note that in $d=2$ we have a relation $w_1^2=w_2$ (see appendix). Note also that $\tF$ is an integral 2-cocycle (twisted or not depending on the symmetry group $G$). There are two independent candidate actions: 
$$
S_1=\frac12 \int_X w_1^2,\quad S_2=\frac12 \int_X \tF.
$$
The action $S_1$ works for both Bosonic Topological Insulators and Bosonic Topological Paramagnets and describes an SPT phase protected solely by $\ZZ_2^T$ \cite{Kcobord}.  On the other hand, $S_2$ defines a nontrivial SPT phase only for $G=U(1)\times\ZZ_2^T$. Indeed, as explained in section 3, for $G=U(1)\rtimes\ZZ_2^T$ the action $S_2$ can be continuously deformed to the trivial action. We conclude that for $D=1$ Bosonic Topological Paramagnets are classified by $\ZZ_2^2$, while Bosonic Topological Insulators are classified by $\ZZ_2$. This agrees with the group cohomology classification \cite{GC}.

Let us examine the two effective actions more closely. The action $S_1$ does not depend on $\tF$ and thus describes a phase which is protected by time-reversal symmetry alone.  It corresponds to a $1D$ system whose endpoints are Kramers doublets, i.e. the boundary degree of freedom has $T^2=-1$. 

The action $S_2$ corresponds to a $1D$ system with symmetry $U(1)\times\ZZ_2^T$ whose endpoints carry half-integral $U(1)$ charge.  This happens  because on a 2d manifold with boundary the action is not invariant: under $\tA\mapsto \tA+\alpha$ where $\alpha$ is an integral 1-cochain, but  transforms as follows:
$$
S_2\mapsto S_2+\frac12 \int_{\partial X} \alpha. 
$$
This can be cured by adding a boundary term $\int A$ with a half-integral coefficient for every component of the boundary. 

The sum of $S_1$ and $S_2$ corresponds to the 1D SPT phase whose endpoints are Kramers doublets and carry half-integral $U(1)$ charge.

\section{$D=2$}

Since the quantum Hall response vanishes for $D=2$, the action must be an integral of a 3-cocycle $L$ constructed out of $w_1,w_2, w_3$ and $\tF$. Note first of all that for any closed 3-manifold ones has $w_1^3=w_1 w_2=w_3=0$ \cite{Thom, MilnorStasheff}, so no nontrivial actions making use only of $\ZZ_2^T$ are possible. (This happens because any 3-manifold is cobordant to the empty set \cite{Thom}). If we allow for $\tF$-dependence, then the only candidate action is
\begin{equation}\label{Sthreed}
S=\frac12 \int w_1\tF.
\end{equation}
However, for $G=U(1)\times \ZZ_2^T$ this action vanishes, see appendix for a proof. Thus we conclude that there are no nontrivial Bosonic Topological Paramagnets for $D=2$. On the other hand, for $G=U(1)\rtimes \ZZ_2^T$ the action (\ref{Sthreed}) is nontrivial, in general. Thus Bosonic Topological Insulators for $D=2$ are classified by $\ZZ_2$. These results agree with \cite{GC,LuVish}. 

Let us examine in more detail the surface state of the nontrivial Bosonic Topological Insulator. Following the approach of \cite{KTlong}, we can construct a topologically ordered surface state by postulating the following lattice 2d action:
$$
S_{2d}=\frac12 \int \left(\phi \delta a+ w_1 a+\phi \tF\right). 
$$
Here $\phi$ is a $\ZZ_2$-valued 0-cochain and $a$ is a $\ZZ_2$-valued 1-cochain.  This action describes a topological $\ZZ_2$-gauge theory coupled to $w_1$ and $A$. When both $w_1$ and $F$ are trivial, the only observables in this theory are Wilson loops $\exp(\pi i \int_\gamma a)$, where $\gamma$ is a closed contour, and vortex operators $\exp(\pi i\phi)$. These observables are topological, i.e. their correlators do not change as one deforms $\gamma$ or the insertion point of the vortex operator. The Wilson loop represents an insertion of a massive particle with a unit $\ZZ_2$ charge.

In general, the equations of motion for $a$ read
$$
\delta\phi=w_1,\quad \delta a= \tF.
$$
The first of these equations means that $\exp(i\pi \phi)$ changes sign as one takes it along an orientation-reversing closed contour. Since $\exp(i\pi \phi)$ is the vortex operator for the gauge field $a$, this means that the vortex operator is odd under time-reversal. The second equation means that the particle which has a unit $\ZZ_2$ charge also has a half-integral $U(1)$ charge. Indeed, to ensure symmetry under a transformation $\tA\ra\tA+\alpha$, where $\alpha$ is an integral 1-cochain, we must postulate $a\mapsto a+\alpha$ modulo $2$. But then the Wilson loop $\exp(\pi i\int_\gamma a)$ is not invariant. One can rectify this by defining an improved Wilson loop
$$
W_{n/2}=\exp(\pi i \int_\gamma (a- n A)),
$$
where $n$ is an odd integer. Such a Wilson loop describes an insertion of a massive particle which has a unit $\ZZ_2$ charge and $U(1)$ charge $n/2$.

\section{$D=3$}

Since the quantum Hall response vanishes for $D=3$, the action must be an integral of a 4-cocycle $L$ constructed out of $w_1,w_2, w_3, w_4$ and $\tF$. It is well-known that for any closed 4-manifold one has $w_1 w_2=0,$ $w_1 w_3=0,$ and $w_1^4+w_2^2+w_4=0$ \cite{Thom, MilnorStasheff}.  Thus we have four independent candidate actions:
$$
S_1=\frac12 \int_X w_1^4,\quad S_2=\frac12 \int_X w_2^2,\quad S_3=\frac12 \int \tF^2,\quad S_4=\frac12\int_X w_1^2 \tF.
$$
Naively, one also has 
$$
S_5=\frac12 \int_X w_2 \tF,
$$
but one can show that this is the same as $S_3-S_4$ (see appendix).  Thus it appears that for $D=3$ we have a $\ZZ_2^4$ classification of both Bosonic Topological Insulators and Bosonic Topological Paramagnets. However, it turns out that for somewhat subtle reasons the action $S_4$ vanishes identically for $G=U(1)\rtimes\ZZ_2^T$. This is also shown in the appendix. Thus 3D Bosonic Topological Insulators  have $\ZZ_2^3$ classification.

The actions $S_1$ and $S_2$ do not depend on $\tF$ and therefore describe SPT phases protected by time-reversal symmetry alone. Topologically ordered surface phases corresponding to these actions have been discussed in \cite{Kcobord} and \cite{Ryan}. A possible surface phase for $S_1$ is a $\ZZ_2$ gauge theory where both electric and magnetic excitations are Kramers doublets \cite{Kcobord}. A possible surface phase for $S_2$ is a $\ZZ_2$ gauge theory where all excitations are fermions \cite{Ryan}. This agrees with the results of \cite{VishSent}. 

A possible surface phase for $S_3$ is easily constructed using the methods of \cite{KTlong}. It is a $\ZZ_2$ gauge theory where both electric and magnetic excitations carry half-integral $U(1)$ charge. The corresponding lattice 3d action is
$$
S_{3d}=\frac12 \int_{\partial X} \left(a\delta b+a \tF+b \tF\right),
$$
where $a$ and $b$ and $\ZZ_2$-valued 1-cochains. This surface phase was also found in \cite{VishSent}. 

In the case $G=U(1)\times\ZZ_2$ we have an additional SPT phase described by the action $S_4$. A possible surface phase is a $\ZZ_2$ gauge theory where electric excitations are Kramers doublets while magnetic excitations carry half-integral $U(1)$ charge. The corresponding lattice 3d action is
$$
S_{3d}=\frac12 \int_{\partial X} \left(a\delta b+a \tF+b w_1^2 \right),
$$

Naively, one can write such an action also for $G=U(1)\rtimes \ZZ_2^T$. However, it turns out that in this case the action can be made  invariant under all symmetries by adding a 3d counterterm which depends only on $\tA$ and the geometry of $\partial X$. This follows from the results of \cite{KTlong} and the fact (proved in the appendix) that the \modtwo\ cohomology class of $w_1^2 F$ is trivial if $F$ is a \modtwo\ reduction of $\tF=\delta_w\tA$. 

Group cohomology classification gives $\ZZ_2^2$ for Bosonic Topological Insulators and $\ZZ_2^3$ for Bosonic Topological Paramagnets. In both cases it misses the SPT phase described by the action $S_2$.

\section{$D=4$}

The last case we consider is $D=4$. In this dimension quantum Hall response is possible for $G=U(1)\rtimes\ZZ_2^T$, since we can write down a Chern-Simons action
$$
S_{CS}=n \int_X \tA \tF^2, \quad n\in\ZZ,
$$
which obeys all the requirements. A mixed $U(1)$-gravitational Chern-Simons term (\ref{CSmixed}) is also possible\footnote{We are grateful to X.-G. Wen for pointing this out. } for $G=U(1)\rtimes\ZZ_2^T$.

All other terms in the action must be constructed out of $w_1,\ldots,w_5$ and $F$. One can show that $w_1 w_2=0$, $w_1 w_3=0$, $w_4+w_1^4+w_2^2=0$, $w_1 w_4=0$, $w_5=0$  for any closed 5-manifold. These relations imply also $w_1^5=0$, so the candidate actions are
$$
S_1=\frac12 \int_X w_2 w_3, \quad S_2=\frac12 \int_X w_3 \tF, \quad S_3=\frac12 \int_X w_1 \tF^2,\quad S_4=\frac12 \int_X w_1^3 \tF .
$$
One can show that the actions $S_2$ and $S_3$ vanish identically both for $G=U(1)\times \ZZ_2^T$ and $G=U(1)\rtimes\ZZ_2^T$ (see appendix). One can also show that the action $S_4$ vanishes identically for $G=U(1)\times\ZZ_2^T$ (see appendix). The action $S_1$ is nontrivial and independent of the $U(1)$ gauge field. In fact, it is nontrivial even on orientable 5-manifolds and is the generator of the oriented cobordism group $\Omega^5_{SO}(pt,U(1))\simeq \ZZ_2$ \cite{Kcobord}.  

We conclude that 4D Bosonic Topological Insulators are classified by $\ZZ\times\ZZ\times \ZZ_2^2$, where one $\ZZ$ factor corresponds to the quantum Hall response and the other one to the mixed electric-thermal Hall response. Bosonic Topological Paramagnets have a $\ZZ_2$ classification (and vanishing quantum Hall response). On the other hand, group cohomology classification gives $\ZZ\times\ZZ_2$ in the former case and a trivial result in the latter case \cite{GC}. It appears to miss the SPT phases described by the action $S_1$, since it involves higher Stiefel-Whitney classes. The action $S_1$ describes a bosonic SPT phase which exists even if all symmetry is broken \cite{Kcobord}. Group cohomology classification also misses the mixed electric-thermal Hall response. 

Let us construct topologically-ordered surface phases of the $D=4$ SPT phases with vanishing quantum Hall response. For the action $S_1$, a possible surface phase has been described in \cite{Kcobord,Ryan}. It is a 4d  topological $\ZZ_2$ gauge theory with an action
$$
S_{4d}=\frac12 \int_{\partial X} \left(b\delta a+w_3 a+w_2 b\right),
$$
where $a$ is a $\ZZ_2$-valued 1-cochain and $b$ is a $\ZZ_2$-valued 2-cochain. Apart from the last two terms describing coupling to background geometry, this is a $\ZZ_2$ BF action, with $a$ playing the role of the gauge field and $b$ being the Lagrange multiplier enforcing a flatness constraint on $a$. 
Such a theory contains electrically-charged particles and magnetically-charged vortex strings. The surface phase corresponding to $S_1$ is such that the electric excitation is a fermion. The vortex string is also fermionic, in the sense that defining a string-insertion operator (i.e. Wilson surface observable) requires a framing of its normal bundle, see \cite{Ryan} for details.

For the action $S_4$ (which is nontrivial only for $G=U(1)\rtimes\ZZ_2^T$) a possible surface phase is again a 4d topological $\ZZ_2$ gauge theory with a coupling to $\tF$ and $w_1$:
$$
S_{4d}=\frac12 \int_{\partial X} \left(b\delta a+w_1^3 a+\tF b\right). 
$$
In such a theory electric excitations carry half-integral $U(1)$ charge, while vortex strings couple to background geometry in a nontrivial way. It would be interesting to understand the physical meaning of this coupling.

\section*{Appendix A}

\subsection*{A.1. Simplicial calculus}

When doing field theory on a lattice, it is often more convenient from a theoretical standpoint to use a triangulation instead of a hypercubic lattice. Triangulations are better than hypercubic lattices because there is a good simplicial analog of the calculus of differential forms, while as far as we know there is nothing similar for hypercubic lattices.

An analog of a $p$-form is a simplicial $p$-cochain, i.e. a function on $p$-simplices.  The space of $p$-cochains on a triangulation $K$ with values in an Abelian group $G$ will be denoted $C^p(K,G)$.  An analog of the exterior differential is the simplicial differential $\delta: C^p(K,G)\ra C^{p+1}(K,G)$. An explicit formula for $\delta$ is \cite{Hatcher}:
\begin{equation}\label{simpdelta}
(\delta f)(v_0\ldots v_{p+1})=\sum_{i=0}^{p+1} (-1)^i f(v_0\ldots \hat v_i\ldots v_{p+1}),
\end{equation}
where $\hat v_i$ means that this argument is not present.  In (\ref{simpdelta}) we assumed that the vertices of $K$ have been ordered in some way,  and $v_0<\ldots <v_{p+1}$ are vertices of a $(p+1)$-simplex. We also used the fact that any $q+1$ vertices of a $p$-simplex, $q<p$, span a $q$-simplex. The simplicial differential satisfies the identity $\delta^2=0$, as usual.

If $G$ is a commutative ring (like $\ZZ$ or $\ZZ_n$), we have an analog of the wedge product: the cup product $\cup $. An explicit formula for $f\cup g$ is \cite{Hatcher}
\begin{equation}\label{cupproduct}
(f\cup g)(v_0\ldots v_{p+q})=f(v_0\ldots v_p) g(v_p \ldots v_{p+q}),
\end{equation}
where $f\in C^p(K,G)$ and $g\in C^q(K,G)$ and it is assumed that $v_0<\ldots < v_{p+q}$. The cup product is associative.

$\delta$ satisfies the usual Leibniz identity with respect to $\cup$:
\begin{equation}\label{Leibniz}
\delta(f\cup g)=\delta f\cup g+(-1)^p f\cup\delta g~.
\end{equation}
The cup product is actually defined in a slightly more general case, when $f\in C^p(K,G)$, $g\in C^q(K,H)$, and there is a bilinear map $G\times H\ra J$ into a third Abelian group $J$. Then we have a cup product $C^p(K,G)\times C^q(K,H)\ra C^{p+q}(K,J)$. The Leibniz identity still holds.
For example, if $G=\ZZ$ and $H$ is an arbitrary abelian group, we have an action of $\ZZ$ on $H$, i.e. a bilinear map $\ZZ\times H\ra H$. Then the cup product makes $C^*(K,H)$ into a graded module over the graded algebra $C^*(K,\ZZ)$. 

Let us go back to the case when $G$ is a ring. Where the simplicial calculus differs from the calculus of forms is in the lack of supercommutativity of the cup product on the cochain level. The cup product of cochains fails to be supercommutative in a very specific way \cite{Steenrod}, so that the cup product of cocycles is supercommutative up to exact terms. Thus the cup product is supercommutative on the level of cohomology classes. The failure of supercommutativity on the cochain level plays an important role in the definition of Steenrod squares \cite{Steenrod} and the Pontryagin square \cite{Whitehead}. 

We will also need to deal with twisted cochains, i.e. cochains valued in a local system of groups \cite{Hatcher}. A local system of groups is a flat bundle of abelian groups. Let $G$ be an abelian group which is a generic fiber of this bundle. For definiteness, we will specialize to the case when the local system is associated to a principal $\ZZ_2$-bundle, and the generator of $\ZZ_2$ acts on $g\in G$ by $g\ra -g$. Principal $\ZZ_2$ bundles are characterized by an element $[w]\in H^1(K,\ZZ_2)$; we will choose a particular 1-cochain $w\in C^1(K,\ZZ_2)$ representing it.  

The main difference between twisted cochains and ordinary cochains is that the value of the twisted cochain depends on the choice of a reference point in the $p$-simplex. It is natural to choose the reference point to be one of the vertices of the simplex. Changing the reference point from $v_i$ to $v_j$ multiplies the value of the twisted cochain by $(-1)^{w(v_i v_j)}$. This is ``parallel transport'' with respect to the ``connection'' $w$. For a $p$-simplex with vertices $v_0<v_1<\ldots < v_p$ it is convenient to always choose $v_0$ as the reference point.  The differential $\delta_w$ for twisted cochains is defined in essentially the same way as for ordinary cochains, but with additional factors which arise from ``parallel transport'':
\begin{equation}
(\delta_w f)(v_0\ldots v_{p+1})=(-1)^{w(v_0 v_1)} f(v_1\ldots v_{p+1})+\sum_{i=1}^{p+1} (-1)^i f(v_0\ldots \hat v_i\ldots v_{p+1}),
\end{equation}
The factor $(-1)^{w(v_0 v_1)}$ arises because the reference point for $f(v_1\ldots v_{p+1})$ is $v_1$ rather than $v_0$, and in order to add it to the rest of the terms one needs to change the reference point to $v_0$. 

We will only use a special case $p=1$, in which case the formula becomes
$$
(\delta_w f)_{v_0 v_1 v_2}=(-1)^{w(v_0 v_1)} f(v_1 v_2)-f(v_0 v_2)+f(v_0 v_1). 
$$
The twisted differential satisfies $\delta_w^2=0$, as usual, and its cohomology is known as cohomology with twisted coefficients. 

\subsection*{A.2. Bockstein homomorphisms}

Let $G_2$ be an abelian group, $G_1$ be its subgroup, and let $G_3=G_2/G_1$. That is, suppose $G_1\ra G_2\ra G_3$ is a short exact sequence of abelian groups. Then we get a long exact sequence in cohomology \cite{Hatcher}:
$$
\ldots \ra H^p(K,G_1)\ra H^p(K,G_2)\ra H^p(K,G_3)\ra H^{p+1}(K,G_1)\ra\ldots
$$
In particular, the homomorphism $H^p(K,G_3)\ra H^{p+1}(K,G_1)$ is known as a Bockstein homomorphism. 

We are interested in several special cases. The first one is $G_1=\ZZ, G_2=\RR$ and $G_3=\RR/\ZZ$, and $p=1$. In this case the Bockstein homomorphism maps a 1-cocycle $A$ with values in $\RR/\ZZ$ to the 2-cocycle $\tF=\delta \tA$, where $\tA$ is a 1-cochain with values in $\RR$ which is a lift of $A$.  

Another special case is $G_2=\ZZ$, $G_1\subset G_2$ the subgroup of even integers, and $G_3=\ZZ_2$. The map $H^p(K,\ZZ_2)\ra H^{p+1}(K,\ZZ)$ will be denoted $\beta$. It is clear that the image of $\beta$ contains only elements of order $2$. One can interpret $\beta(a)$ as an obstruction for lifting a class $a\in H^p(K,\ZZ_2)$ to a class in $H^p(K,\ZZ)$.

The last case is $G_1=\ZZ_2, G_2=\ZZ_4, G_3=\ZZ_2$. The map  $H^p(K,\ZZ_2)\ra H^{p+1}(K,\ZZ_2)$ will be denoted $\beta_2$. It is clear that $\beta_2$ is a composition of $\beta$ and reduction modulo $2$. This implies that $\beta_2\circ\beta_2=0$. It also implies that $\beta_2(a)$ vanishes for any $a$ which is a reduction modulo $2$ of an integral class. Furthermore, it is easy to show that $\beta_2$ satisfies the Leibniz rule, i.e. for any $f\in H^p(K,\ZZ_2)$ and $g\in H^q(K,\ZZ_2)$ we have
$$
\beta_2 (f\cup g)=\beta_2(f)\cup g+f \cup \beta_2(g). 
$$
Thus $\beta_2$ is a differential on the graded commutative algebra $H^*(K,\ZZ_2)$.

\subsection*{A.3. Steenrod squares and Stiefel-Whitney classes}

For any topological space $X$ and any $i> 0$, Steenrod squares are homomorphisms $Sq^i: H^p(X,\ZZ_2)\ra H^{p+i}(X,\ZZ_2)$. These operations are ``natural'', i.e. behave well under continuous maps of topological spaces and can be defined axiomatically \cite{MilnorStasheff}, as well as by an explicit construction \cite{Steenrod}.  The most important properties for us are:
\begin{itemize}

\item If $a\in H^i(X,\ZZ_2)$, then $Sq^i a=a\cup a$.

\item For $i=1$ and $a\in H^p(X,\ZZ_2)$ we have $Sq^1 a=\beta_2(a)$, where $\beta_2: H^p(X,\ZZ_2)\ra H^{p+1}(X,\ZZ_2)$ is the Bockstein homomorphism. 

\end{itemize} 
These properties imply, in particular, that for any $a\in H^1(X,\ZZ_2)$ we have $\beta_2(a)=Sq^1a=a\cup a$. This fact is used below for $a=w_1$. 

Now let $X$ be a closed $d$-manifold. Then one can define Stiefel-Whitney classes $w_p(X)\in H^p(X,\ZZ_2)$, $p=0,1,\ldots, d$ \cite{MilnorStasheff}. There are important relations between Stiefel-Whitney classes and Steenrod squares. In particular, if $a\in H^{d-i}(X,\ZZ_2)$, then $Sq^i a=v_i(X) \cup a$, where $v_i(X)\in H^i(X,\ZZ_2)$ is a certain polynomial in Stiefel-Whitney classes. The expression for $v_i(X)$ does not depend on $X$. The classes $v_0(X),\ldots,v_d(X)$ are called Wu classes. The first few Wu classes are
$$
v_0=1,\quad v_1=w_1,\quad v_2=w_1^2+w_2,\quad  v_3=w_1 w_2.
$$
For example, for $d=4$ and any $a\in H^2(X,\ZZ_2)$ we have
$$
Sq^2 a=(w_1^2+w_2) \cup a.
$$
Using the first property of Steenrod squares, we get an identity which holds for any $a\in H^2(X,\ZZ_2)$ on any closed 4-manifold $X$:
$$
(w_1^2+w_2)\cup a=a\cup a. 
$$
In particular, this shows that the action $S_5$ in section 7 is the differences of $S_3$ and $S_4$. 

Similarly, for any closed 3-manifold $X$ and any $a\in H^2(X,\ZZ_2)$ we have
$$
w_1\cup a=\beta_2(a). 
$$
If $a$ is a reduction of an integral cohomology class, then $\beta_2(a)=0$. This proves that on any closed 3-manifold one has $w_1 F=0$, where $F$ is a reduction modulo two of $\tF=\delta \tA$, where $A\in C^1(X,\RR)$ is closed modulo integers. This fact is used in section 6 to show that the action (\ref{Sthreed}) is trivial for $G=U(1)\times\ZZ_2^T$. 

Now let $X$ be a closed 5-manifold and $a$ be a  2-cocycle modulo $2$. Then $w_1 a^2=0$. Indeed, 
$$
w_1 a^2=Sq^1(a\cup a)=Sq^1(a)\cup a+a\cup Sq^1(a)=0.
$$
This implies that the action $S_3$ in section 8 is trivial. 

If in addition $a\in H^2(X,\ZZ_2)$ is a reduction of an integral cocycle $\tilde a$, then $w_1^3 a$ also vanishes. Indeed:
$$
w_1^3 a=Sq^1(w_1^2 a)=Sq^1(Sq^1(w_1)\cup  a)=Sq^1(w_1)\cup Sq^1(a)=w_1^2\cdot 0=0.
$$
This implies that the action $S_4$ in section 8 is trivial for $G=U(1)\times\ZZ_2^T$. Similarly, using the fact that $w_3=Sq^1 w_2$, we get that $w_3 a$ also vanishes:
$$
w_3 a=Sq^1(w_2) a=Sq^1(w_2 a)=w_1 w_2 a=0.
$$
This implies that the action $S_2$ in section 8 is trivial for $G=U(1)\times\ZZ_2^T$.

We also claimed that the action $S_4$ in section 7 and the action $S_2$ in section 8 vanish for $G=U(1)\rtimes\ZZ_2^T$. This is less trivial. We need the following lemma. Let $X$ be a triangulated topological space, $w$ be a 1-cocycle with values in $\ZZ_2$, and let $\ZZ^T$ and $\RR^T$ be local systems on $X$ with fibers 
$\ZZ$ and $\RR$ twisted by $w$. Further, let  $\tA$ be a twisted 1-cochain on $X$ valued in $\RR^T$ such that $\tF=\delta_w\tA$ is a twisted 2-cochain valued in $\ZZ^T$, and let   
$F\in H^2(X,\ZZ_2)$ be the reduction of $\tF$ modulo $2$. Then $\beta_2(F)=w\cup F$.

The proof is a straightforward computation using simplicial cochains. The lemma itself can be motivated as follows. If $w$ is trivial, then $\tF$ is an integral cohomology class, hence $\beta_2(F)=0$. This suggests that for general $w$ the class $\beta_2(F)\in H^3(X,\ZZ_2)$ is not independent and should be expressible through $w$ and $F$. It also must vanish when either $w=0$ or $F=0$. This leaves $w\cup F$ as the only candidate.

Given this lemma, it is easy to prove the desired results. If $X$ is a closed 4-manifold, $w=w_1(X)$, and $\tF$ and $F$ are as above, then 
$$
w_1^2F=Sq^1(w_1\cup F)=Sq^1(Sq^1(F))=0.
$$ 
This implies that the action $S_4$ in section 7 is trivial for Bosonic Topological Insulators. Similarly, if $X$ is a closed 5-manifold, then
$$
w_3 F=Sq^1(w_2)F= Sq^1(w_2 F)+w_2 Sq^1 F=w_1 w_2 F+w_2 w_1 F=0.
$$
This implies that the action $S_2$ in section 8 vanishes for Bosonic Topological Insulators.

In this paper we often made  use of various algebraic relations between Stiefel-Whitney classes \cite{Thom, MilnorStasheff}. These relations depend on the dimension of the manifold $X$ and can be derived as follows. First, Steenrod squares satisfy $Sq^i a=0$ if $a\in H^p(X,\ZZ_2)$ and $i>p$. This implies that for a closed $d$-manifold the Wu class $v_p(X)$ vanishes if $p>d/2$. Consequently all its Steenrod squares also vanish. This gives all the relations between Stiefel-Whitney classes in dimension $d$. To compute Steenrod squares of Wu classes one uses other properties of Stiefel-Whitney classes and Steenrod squares, such as the Wu formula and the Cartan formula \cite{Thom, MilnorStasheff}.

\end{document}